\documentclass[%
 amsmath,amssymb,
preprint,%
]{revtex4-1}

\usepackage{graphicx}% Include figure files
\usepackage{dcolumn}% Align table columns on decimal point
\usepackage{bm}% bold math
\usepackage[utf8]{inputenc}
\usepackage[T1]{fontenc}
\usepackage{mathptmx}
\usepackage{etoolbox}
\usepackage{siunitx}

\usepackage{titlesec}
\titlespacing*{\section}{0pt}{10pt}{6pt}
\usepackage{parskip}
\newcommand{\editorr}[2]{%
  \expandafter\newcommand\csname #1note\endcsname[1]{%
    \textcolor{#2}{(\textbf{#1:} ##1)}}%
  \expandafter\newcommand\csname #1\endcsname[1]{%
    \textcolor{#2}{##1}}%
  \expandafter\newcommand\csname #1cancel\endcsname[1]{%
    \textcolor{#2}{\sout{##1}}}%
  \expandafter\newcommand\csname #1change\endcsname[2]{%
    \textcolor{#2}{\sout{##1} ##2}}%
  \newenvironment{#1text}{\color{#2}}{\color{black}}
}

\makeatletter
\def\@email#1#2{%
 \endgroup
 \patchcmd{\titleblock@produce}
  {\frontmatter@RRAPformat}
  {\frontmatter@RRAPformat{\produce@RRAP{*#1\href{mailto:#2}{#2}}}\frontmatter@RRAPformat}
  {}{}
}%
\makeatother
\begin{document}

\title{Experimentally constrained modeling of the Pockels response of KNbO$_3$ and KTaNbO$_3$}

% Force line breaks with \\
\author{Virginie de Mestral}
 \email{vdemestral@ethz.ch}
 \altaffiliation{Integrated Systems Laboratory, ETH Zurich, Zurich, Switzerland}

\author{Lorenzo Bastonero}
 \altaffiliation{U Bremen Excellence Chair, Bremen Center for Computational Materials Science, MAPEX Center for Materials and Processes, University of Bremen, D-28359 Bremen, Germany}

\author{Petr Bednyakov}
 \altaffiliation{Institute of Physics, Czech Academy of Sciences, Prague, Czech Republic}

 \author{Fedir Borodavka}
 \altaffiliation{Institute of Physics, Czech Academy of Sciences, Prague, Czech Republic}

\author{Simon Mellaerts}
\altaffiliation{Integrated Systems Laboratory, ETH Zurich, Zurich, Switzerland}

\author{Tetyana Ostapchuk}
 \altaffiliation{Institute of Physics, Czech Academy of Sciences, Prague, Czech Republic}

\author{Alex Pescaru}
 \altaffiliation{Institute of Physics, Czech Academy of Sciences, Prague, Czech Republic}

\author{Jiri Hlinka}
  \altaffiliation{Institute of Physics, Czech Academy of Sciences, Prague, Czech Republic}

\author{Mathieu Luisier}
  \altaffiliation{Integrated Systems Laboratory, ETH Zurich, Zurich, Switzerland}

\date{\today}% It is always \today, today,
             %  but any date may be explicitly specified

\begin{abstract}
The soft-mode of electro-optic (EO) metal-oxide perovskites plays a critical role in determining their Pockels responses. This is the case of potassium tantalate niobate (KTN), which exhibits an intrinsic Pockels response $2.5$ times larger than that of state-of-the-art barium titanate (BTO), highlighting its potential for high-performance EO applications. By combining \textit{ab initio} calculations at the density-functional theory (DFT) level and far-IR measurements, we reveal that the harmonic approximation combined with semi-local exchange-correlation functionals fails to accurately capture the soft transverse optical (TO) Slater mode of both potassium niobate (KNO) and KTN, which dominates the technologically relevant $r_{51}$ Pockels coefficient. Replacing the calculated mode frequency with its measured value provides an experimentally constrained approach that substantially improves the predicted Pockels response. For that purpose, the previously unreported TO Slater-mode frequency of KTN is extracted from far-IR reflectivity measurements. The results emphasize the potential of KTN-based EO modulators as an alternative to standard lithium niobate (LNO) and BTO technologies, with a potentially lower energy consumption and device footprint. 
\end{abstract}

\maketitle

The data centers driving today’s AI revolution rely heavily on graphics processing units (GPUs) for efficient massive parallel computations. High-bandwidth interconnects are therefore required to exchange data between GPUs and across large-scale clusters~\cite{demkov2024conference}. To support current and future workloads, traditional electrical links are rapidly being replaced by electro-optic (EO) interconnects. Co-packaged optics (CPO) on monolithic silicon platforms is a leading solution for low-loss, high-bandwidth inter-GPU communication~\cite{dourado2021figure_of_merit}. Although dense photonic integrated circuits (PICs) can mitigate latency, cross-talk, and dispersion, EO modulators remain a severe bottleneck due to limited speed, high driving voltages, and large footprints~\cite{margalit2021Siphotonics_vs_electronics}. Efficient phase shifting can be achieved by exploiting the Pockels effect in non-centrosymmetric crystals integrated into a Mach-Zehnder interferometer (MZI)~\cite{abel2019largePockels}, where the refractive index varies linearly with an applied electric field. Since the MZI length and driving voltage are inversely proportional to the Pockels response, maximizing the latter is essential to reduce optical insertion losses and device footprints.

Lithium niobate (LNO) is the industry standard, with a Pockels coefficient~\cite{chmielak2011pockels_strainedSi} $r_{33}$ of about $30$ \unit{\pico\meter\per\volt}. Recent advances in barium titanate (BTO) have demonstrated excellent crystal quality, silicon compatibility, and a remarkable $r_{51}$ of $923$ \unit{\pico\meter\per\volt} in thin films\cite{abel2019largePockels}. Potassium niobate tantalate (KTN) even exceeds this mark with the largest Pockels coefficient reported to date~\cite{loheide1993KTN_r33,vanRaalte1967linearEO_KTN,haas1967linearEO_KTN, neumann1999linearEO_KTN_BCT}. This solid solution of potassium niobate (KNO) and potassium tantalate (KTO)~\cite{shang2022giant_eo_ktn_domain} is expected to reach $51200$ \unit{\pico\meter\per\volt}.

KNO is a perovskite with a Curie temperature of $T_c=708$ K. It is tetragonal (\textit{P4mm}) below $T_c$, becomes orthorhombic (\textit{Amm2}) at $498$ K, and rhombohedral (\textit{R3m}) at $228$ K~\cite{yelon1971neutron_scatt_soft_mode_KTN}. Above $T_c$, it is cubic and centrosymmetric, thus showing no Pockels response. The \textit{Amm2} phase is stable at room temperature and has a clamped $r_{51}$ of $360$ \unit{\pico\meter\per\volt}. As in BTO, its phase transitions are driven by a Slater-type soft TO mode, in which the anions and cations oscillate in opposite directions. Consequently, the polarization changes from $[001]$ in \textit{P4mm}, to $[011]$ in \textit{Amm2}, and finally to $[111]$ in \textit{R3m}~\cite{hewat1973KNO_phases_transition}. In contrast, KTO is an incipient ferroelectric whose phase transitions are suppressed by quantum fluctuations below $4$ K~\cite{esswein2022ferroelectric_paraelectric}; It is therefore centrosymmetric and does not have a Pockels response.

Mixing KNO and KTO yields KTN, which exhibits the same phase transitions and soft-mode characteristics as KNO for niobium concentrations above $5\%$~\cite{triebwasser1959phase_transition_KTN}. The Pockels response of these compounds can be exceptionally large. In the tetragonal (\textit{P4$2$nm}) phase, $r{33}$ ranges from $240$ to $1750$ \unit{\pico\meter\per\volt} when the niobium concentration decreases from $52$\% to $43$\%~\cite{loheide1993KTN_r33}. The $r_{51}$ coefficient reaches $5770$--$7850$ \unit{\pico\meter\per\volt} for $47$\%--$52$\% niobium~\cite{neumann1999linearEO_KTN_BCT}, and can be further increased to $51200$ \unit{\pico\meter\per\volt} through ferroelectric domain engineering~\cite{shang2022giant_eo_ktn_domain}. These high Pockels coefficients offer a promising route to next-generation EO modulators with enhanced performance compared to current LNO-based devices. High-quality single-phase KTN thin films can also be monolithically integrated onto silicon~\cite{guiller2008KTN_on_Si}.

In this work, we focus on tetragonal, equi-partitioned KTa$_{0.5}$Nb$_{0.5}$O$_3$ (KTN50), which is stable at room temperature and can be modeled using the fewest atoms possible. We consider its clamped Pockels response, as the thin film is assumed to be embedded in a silicon cladding. We first validate our modeling approach against the room-temperature-stable \textit{Amm2} phase of KNO, for which experimental Pockels coefficients are available. We then estimate the Pockels tensor of KTN by computing its TO Slater phonon mode value and replacing it with its experimental value extracted from far-IR reflectivity measurements.

Within the harmonic approximation to crystal vibrations~\cite{MaxBorn1954}, the $r_{ijk}$ entries of the clamped Pockels tensor can be calculated according to the formalism of Veithen \textit{et al.}~\cite{veithen2005nonlinear_DFPT}
\begin{equation}
    r_{ijk} = -\frac{2}{n_i^2 n_j^2}\chi_{ijk}^{(2)}
    -\frac{1}{n_i^2 n_j^2} \sum_m \frac{\alpha_{ij}^m p_{m,k}}{\omega_m^2},
    \label{equ:method:pockels_Veithen}
\end{equation}
where $\chi_{ijk}^{(2)}$ is the second-order dielectric susceptibility, $n_{i}$ the refractive index, $\alpha_{ij}^m$ the Raman susceptibility, $p_{m,k}$ the mode polarity (oscillator strength), and $\omega_m$ the frequency of the $m^{\mathrm{th}}$ phonon mode at the $\Gamma$-point. The $i$, $j$, and $k$ indices refer to the polarization of the incoming and outgoing light waves, as well as of the modulating electric field, respectively. 

\begin{figure}[h!]
    \centering
    \includegraphics[width=0.5\linewidth]{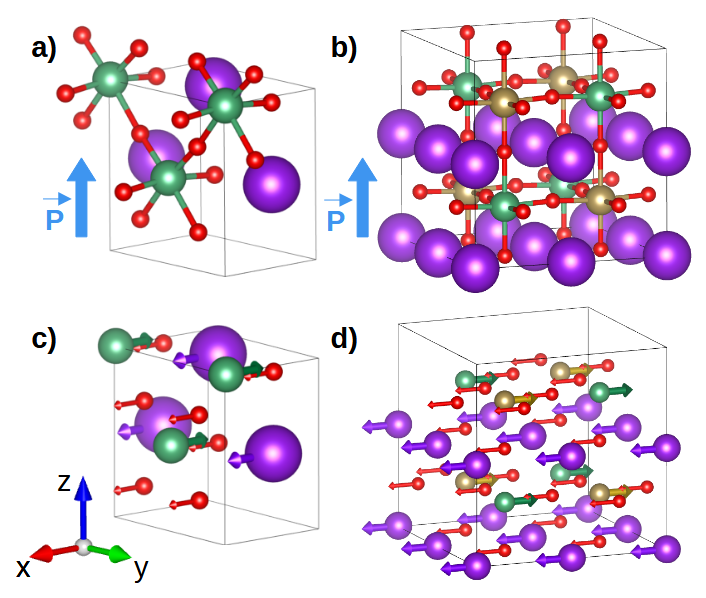}
    \caption{Representation of a) the conventional cell of \textit{Amm2} KNO and b) the supercell of \textit{P4$_{2}$nm} KTN50 with potassium in purple, niobium in green, tantalum in gold, and oxygen in red. The eigenvectors associated with the soft TO Slater mode of each structure are plotted in c) and d), respectively.}
    \label{fig:intro_structures}
\end{figure}

We compute the $r_{ijk}$ tensor at the \textit{ab initio} level with the \texttt{aiida-vibroscopy} package~\cite{bastonero2024Vibroscopy}, as described in Ref.~\cite{demestral_pockels_2025}, specifically employing the \textit{IRamanSpectraWorkChain} module. All DFT calculations rely on the Perdew-Burke-Ernzerhof XC-functional for solids (PBEsol)~\cite{perdew1996GGA} combined with scalar relativistic, projector augmented-wave (PAW) pseudopotentials using a $4s^14p^03s^23p^6$ valence configuration for potassium, $5s^25p^04s^24p^64d^3$ for niobium, $6s^26p^05s^25p^65d^3$ for tantalum, and $2s^22p^4$ for oxygen~\cite{dalCorso2014USPP_generation}. We utilize a conventional 10-atom unit cell to describe the $\textit{Amm2}$ lattice of KNO (see Fig.~\ref{fig:intro_structures} a)) and a ($2\times2\times2$) 5-atom supercell to model the \textit{P4$_{2}$nm} KTN structure, as illustrated in Fig.~\ref{fig:intro_structures} b). 

The optimal lattice parameters are determined with plane-wave and charge density kinetic energy cutoffs of 80 Ry and 800 Ry, respectively. $\Gamma$-centered ($8\times6\times6$) and ($4\times4\times4$) Monkhorst-Pack k-point grids are used for $\textit{Amm2}$ KNO and KTN50, respectively. The forces are converged within $10^{-4}$ Ry/atom and the total energy within $10^{-5}$ Ry/atom. The $\textit{DielectricWorkChain}$ workflow of \texttt{aiida-vibroscopy} is run with a finite electric field step of $5\cdot10^{-4}$ Ry a.u $\approx 0.018$ V/\AA~. The k-point density is multiplied by a factor of $1.25$ along the direction of the applied field~\cite{bastonero2024Vibroscopy}. Finite atomic displacements of 0.01 \AA~ are used when running the $\textit{HarmonicWorkChain}$ workflow, while the force constants are symmetrized through the translational and the permutational acoustic sum rules.

\begin{figure}[h!]
\includegraphics[width=0.7\linewidth]{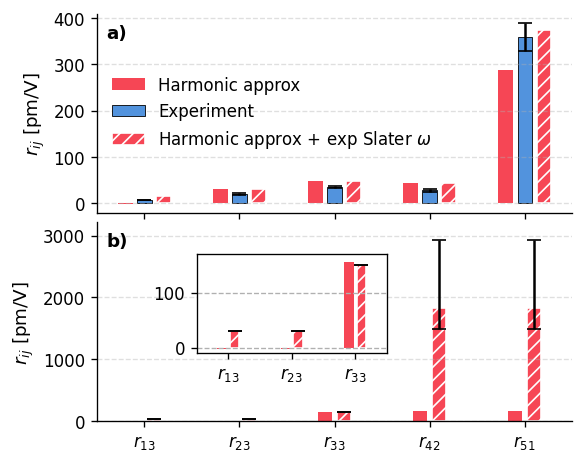}
\caption{Pockels coefficients of a) \textit{Amm2} KNO and b) \textit{P4$_2$nm} KTN50 calculated with \texttt{aiida-vibroscopy} in the harmonic approximation limit (solid red bars) and after replacing the computed Slater mode frequencies by their measured values (dashed red bars). For \textit{Amm2} KNO, the experimental Pockels coefficients are reported as blue bars~\cite{zgonik1993materials_KNO_EO}.}.
\label{fig:res_pockels}
\end{figure}

The properties of the full KTN compositional range have not yet been reported in the literature. In particular, to the best of our knowledge, neither the Slater mode frequency nor the clamped Pockels tensor of KTN50 have been experimentally determined. Hence, our objective is to fill this gap and provide these quantities as accurately as possible based on \textit{ab initio} simulations and far-IR Slater-mode measurements. As a first step, to validate our method, we calculate the Pockels tensor of \textit{Amm2} KNO for which experimental data is available. In its conventional cell coordinate system, \textit{Amm2} KNO has a total polarization aligned with the $[001]$ axis, as shown in Fig.~\ref{fig:intro_structures} a). We find that \textit{Amm2} KNO displays a dynamical instability i.e., an imaginary frequency at the $\Gamma$-point, associated with a soft, $x$-polarized TO Slater mode. The eigenvectors of the latter are plotted in Fig.~\ref{fig:intro_structures} c). This Slater mode is underdamped i.e., perturbed by a weak anharmonicity~\cite{hurrell1975orthorhombic_KNO_overdamped}. To stabilize it within the harmonic approximation, we first relax the atomic positions of the selected unit cell in Fig.~\ref{fig:intro_structures} a) along the path suggested by its Slater eigenvectors. This yields a distorted, dynamically stable KNO structure with total polarization along the $[101]$ axis and a two-fold symmetry around the $[001]$ axis. The geometric average of both symmetrically-equivalent structures preserves the original \textit{Amm2} polarization direction along $[001]$. We apply the same procedure to the evaluation of the Pockels tensor~\cite{kim2023Pockels_tetra_BTO,demestral_pockels_2025}, i.e., we compute a dynamical average of symmetrically-equivalent, $180$\unit{\degree} rotated Pockels tensors to preserve the macroscopic properties associated with the original \textit{Amm2} lattice. The resulting Pockels coefficients are shown as solid red bars in Fig.~\ref{fig:res_pockels} a), where the Voigt notation is used ($r_{113} \rightarrow r_{13}, r_{223} \rightarrow r_{23}, r_{333} \rightarrow r_{33}, r_{232} \rightarrow r_{42}, r_{131} \rightarrow r_{51}$). We observe a general quantitative agreement with experimental data~\cite{zgonik1993materials_KNO_EO}, except for the coefficient of highest technological relevance, $r_{51}$, which underestimates the lower bound of the error bar by 30 \unit{\pico\meter\per\volt} corresponding to a relative error of 20\%.

\begin{figure}[h!]
    \centering
    \includegraphics[width=0.85\linewidth]{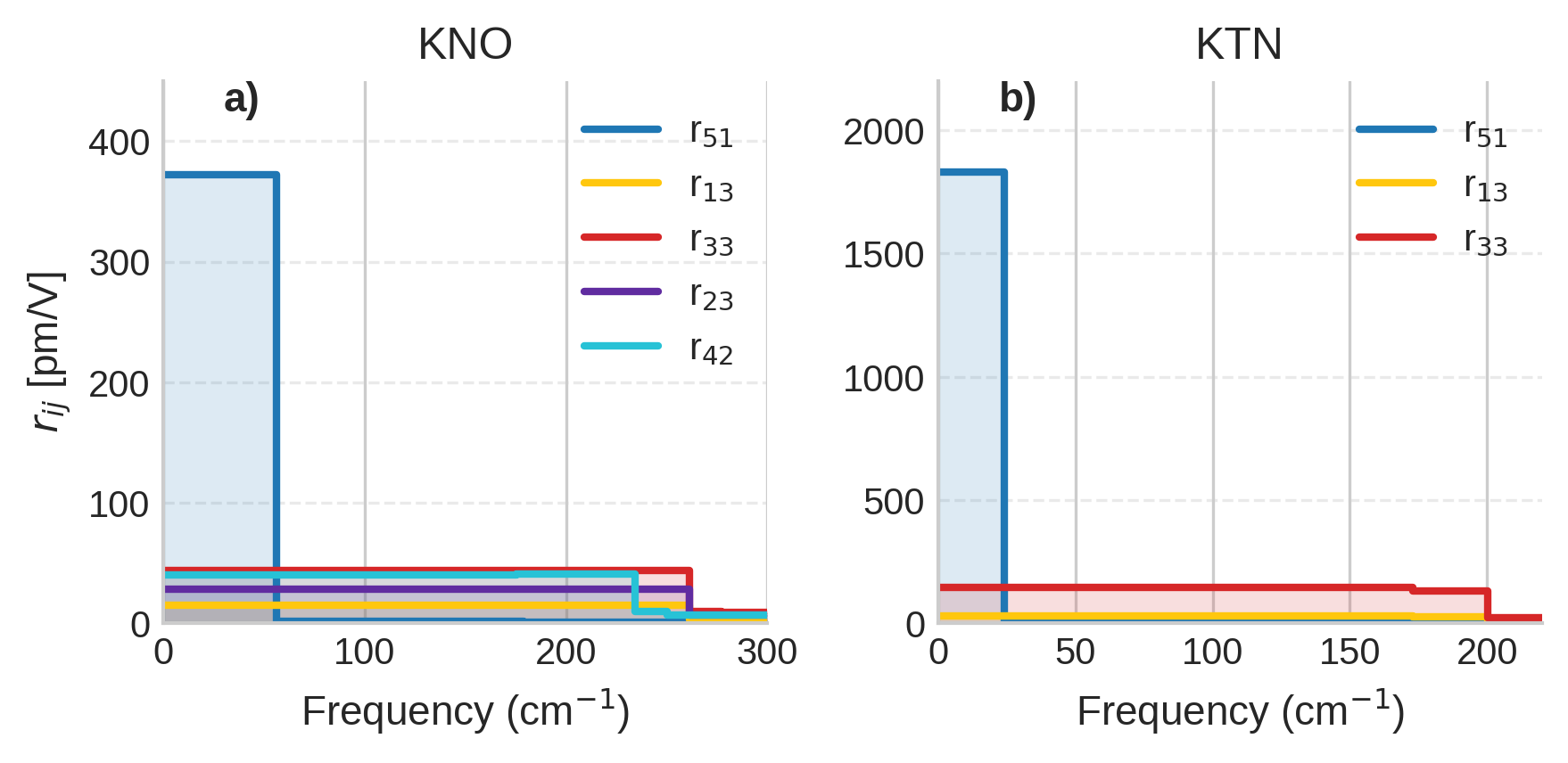}
    \caption{Cumulative, phonon frequency-dependent Pockels coefficients $r_{51}$, $r_{13}$, $r_{33}$, $r_{23}$, and $r_{42}$ of a) \textit{Amm2} KNO and b) \textit{P4$_2$nm} KTN50. The Pockels response is largest at low frequencies and decreases in a step-like fashion with growing frequency, as phonon modes freeze out. In the high-frequency limit, the Pockels response is null.}
    \label{fig:bandwidth}
\end{figure}

To identify the origin of the discrepancy between the predicted and experimental values of $r_{51}$ in \textit{Amm2} KNO, we return to the definition of the Pockels tensor in Eq.~(\ref{equ:method:pockels_Veithen}). It turns out that the factor causing the largest variation in $r_{ijk}$ is $1/\omega^2_m$ and the phonon mode that contributes the most, and by a large margin, to the Pockels response is the Slater one, as can be seen in Fig.~\ref{fig:bandwidth} a). As a consequence, the calculation of $r_{51}$ can be simplified by retaining only this single mode. Since the phonon properties of perovskite oxides strongly depend on both the exchange-correlation (XC) functional and the harmonic approximations~\cite{Verdi2023} and the best parameter setting has not yet been determined, we tried to replace the calculated harmonic Slater mode frequency by the experimental one~\cite{hurrell1975orthorhombic_KNO_overdamped}, $56$ \unit{\per\centi\meter} in our modeling approach. Through this substitution and keeping a single-mode contribution, $r_{51}$ in the high-symmetry phase of KNO increases to $374$ \unit{\pico\meter\per\volt}, which is in good agreement with the experimental Pockels value~\cite{zgonik1993materials_KNO_EO} of $360 \pm 30$ \unit{\pico\meter\per\volt}.

After validating our method with \textit{Amm2} KNO, we apply it to the calculation of the Pockels tensor of tetragonal \textit{P4$_2$nm} KTN50. This compound exhibits two degenerate dynamic instabilities at $\Gamma$, associated with the $x$- and $y$-polarized E(TO) Slater phonon modes illustrated in Fig.~\ref{fig:intro_structures} d). Relaxing the atomic positions along the directions of the combined Slater mode displacements produces a dynamically stable, distorted KTN50 cell with total polarization aligned with $[111]$. Here again, we perform a dynamical average of the clamped Pockels tensors associated with each four-fold symmetry-equivalent representation, preserving the original \textit{P4$_2$nm} symmetry and polarization along $[001]$ of KTN50. Results are shown as solid red bars in Fig.~\ref{fig:res_pockels} b). Despite the demonstrated EO properties of KTN in general, our results indicate that the degenerate $r_{42}$ and $r_{51}$ coefficients of the considered compound do not exceed $162$ \unit{\pico\meter\per\volt}, well below the expected range.

\begin{figure}[h!]
    \centering
    \includegraphics[width=0.6\linewidth]{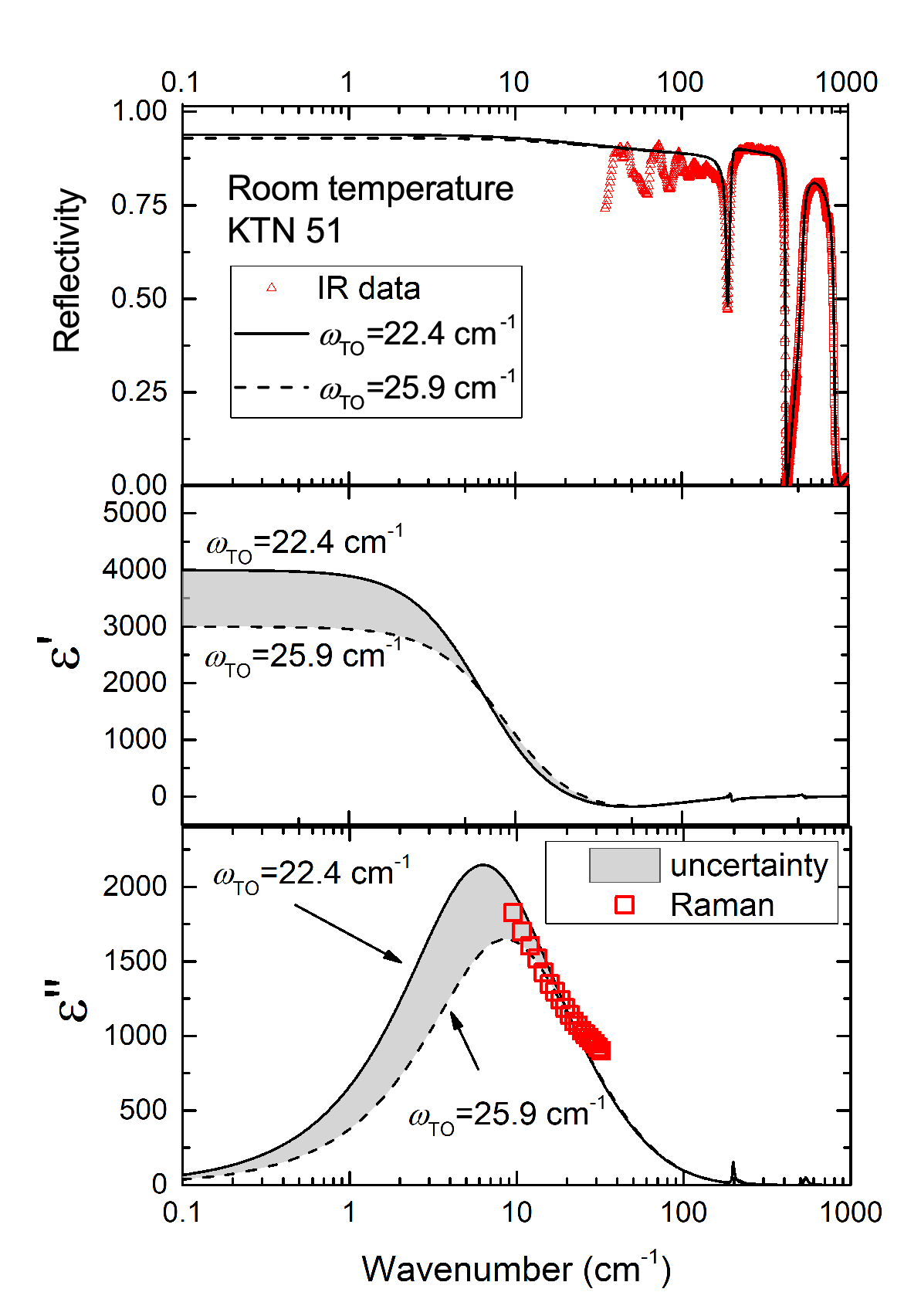}
    \caption{Room-temperature dielectric spectrum of a KTN51 single crystal~\cite{bednyakov2013dielectric}. a) Room-temperature IR reflectivity spectrum fitted to the measurement of a polished crystal plate.  b) Real part of the dielectric permittivity spectrum. c) Corresponding imaginary part. The black lines refer to the results of a damped harmonic oscillator model fitted to the measured (red dots) reflectivity spectrum in a) and Bose-factor-corrected Raman susceptibility in c). The latter was scaled by a multiplicative constant to match the known static dielectric permittivity of KTN51 (3000-4000)~\cite{bednyakov2013dielectric}.}
    \label{fig:experiment}
\end{figure}

As in KNO and shown in Fig.~\ref{fig:bandwidth} b), the Pockels response of KTN50 is entirely dominated by its Slater mode whose frequency in the harmonic limit is equal to $60$ \unit{\per\centi\meter}. Next, we replace this value by the experimental Slater mode frequency to better estimate the $r_{42}$ and $r_{51}$ coefficients. As this parameter is not available in the literature, we measured it by recording the far-IR reflectivity of a KTa$_{0.51}$Nb$_{0.49}$O$_{3}$ (KTN51) single crystal with a composition as close as possible to that of the equipartitioned KTN50 supercell, as detailed in Fig.~\ref{fig:experiment}. The selected sample was previously characterized in Ref.~\onlinecite{bednyakov2013dielectric}, and has a static permittivity in the order of $3000$-$4000$. The optical interference fringes in the reflectivity below $100$ \unit{\per\centi\meter}, arising due to the aperture delimiting the almost single domain area of the crystal, were ignored when fitting the measured spectrum with a damped harmonic oscillator model~\cite{berreman1968DHO}. In addition, the polarized Raman spectrum, which is sensitive to E(TO) modes, was recorded in the same area as the reflectivity measurements. Since the spectral range below $100$ \unit{\per\centi\meter} is largely dominated by the contribution of the Slater mode alone, it can be assumed that the Bose-Einstein factor corrected Raman spectrum is proportional to the dielectric loss spectrum in this frequency range ({\it i.e.} that the scattering and loss spectra of this mode are related by the fluctuation-dissipation theorem~\cite{ondrejkovic2014dynamics}). In this way, both Raman and reflectivity data could be fitted together with an imposed value of the static permittivity. Assembling everything, assuming static permittivity in the order of $3000$-$4000$, the transverse optical mode frequency of the KTN51 Slater mode was determined to be about $24\pm5$ \unit{\per\centi\meter}. As this frequency is significantly lower than our computed value, we suspect that the XC functional approximation along with the harmonic approximation cannot accurately capture the behavior of the soft TO Slater mode of KTN50, which was found experimentally to be overdamped for compositions above $33$\unit{\percent} of niobium~\cite{yelon1971neutron_scatt_soft_mode_KTN,courdille1977overdamped_KTN}.

With a Slater mode frequency of $24\pm5$ \unit{\per\centi\meter}, the $r_{51}$ and $r_{42}$ coefficients of the high-symmetry \textit{P4$_2$nm} phase of KTN50 increase to $\sim1800$ \unit{\pico\meter\per\volt}, as indicated in Fig.~\ref{fig:res_pockels} b), with an upper and lower bound of $2900$ and $1500$ pmV, respectively. The latter are due to the $\pm 5$ \unit{\per\centi\meter} uncertainty in the Slater mode frequency. Notably, even the upper bound is still clearly smaller than the expected $5770$ to $7850$ \unit{\pico\meter\per\volt} range that was measured for the unclamped $r_{51}$ coefficient of KTN50, with $47\%$ to $53\%$ niobium, respectively~\cite{neumann1999linearEO_KTN_BCT}. This significant underestimation is most likely caused by the omission of the piezoelectric contribution and linearized Kerr response of this material. Indeed, coefficients measured with an above-ground sinusoidal voltage signal, as in Ref.~\onlinecite{vanRaalte1967linearEO_KTN}, actually include both a linearized Kerr and Pockels contribution~\cite{steglich2020linearized_Kerr}. This known phenomenon is particularly prominent in KTN due to its large Kerr coefficient~\cite{tadayuki2007linearized_Kerr}.  

Capturing these effects (piezoelectricity and Kerr) and more accurately predicting the EO properties of KTN would require integrating the missing physics into our model. However, in this work, we restrict ourselves to the intrinsic clamped Pockels response whose characteristics should be properly described before going to the next step. Remarkably, the intrinsic $r_{51}$ coefficient of KTN\cite{zgonik1994BTOprop} already exceeds that of BTO ($730$ \unit{\pico\meter\per\volt}) by a factor of $2.5$ thanks to a reduced Slater mode frequency~\cite{ostapchuk2005soft_BTO} $24\pm 5$ vs. $40$ \unit{\per\centi\meter}. This result emphasizes the potential of KTN-based MZI as an alternative to LNO and BTO technologies, with potentially lower driving voltages and circuit footprints.

To conclude with, \textit{ab initio} calculations of the clamped Pockels tensor of KTN severely underestimate the coefficient of highest technological relevance, $r_{51}$ when using the harmonic approximation coupled with semi-local XC functionals. An analysis of the different phonon modes contributing to $r_{51}$ showed that the Slater one, which drives phase transitions, dominates over all the others. As $r_{51}$ scales with $1/\omega^2$, it is very sensitive to this parameter, which should be determined very accurately. By replacing the calculated frequency with the measured one, the experimental Pockels response is better reproduced, but at the expense of the prediction capabilities of our model. Indeed, we had to measure the Slater mode frequency of KTN51 from its far-IR reflectivity spectrum, thus requiring the availability of such a material sample. Doing so gave rise to an intrinsic $r_{51}$ value of $1832$ \unit{\pico\meter\per\volt}, a $2.5\times$ increase as compared to BTO, the current state-of-the-art Pockels compound. 
At the same time, the inverse-square dependence of the Pockels response on the Slater mode frequency provides a powerful route for materials optimization. We illustrated this concept through composition engineering: By alloying KNO with $50\%$ Ta, its Slater mode frequency decreased from $56$ to $24\pm5$ \unit{\per\centi\meter}, which enhanced its $r_{51}$ coefficient by a factor of $5$.
\newline

The data that support the findings of this study are openly available in the Materials Cloud at https://archive.materialscloud.org/record/2026.173, reference number \\ https://doi.org/10.24435/materialscloud:xv-n3.
\newline

The authors have no conflicts to disclose.
\newline

This project was funded by Innosuisse, the Swiss Innovation Agency, under Project No. 122.037 IP-ENG. Computer time was provided by the Swiss National Supercomputing Centre (CSCS) under projects lp16 and lp82. Funding was also provided by the Ferroic Multifunctionalities project, supported by the Ministry of Education, Youth, and Sports of the Czech Republic, Project No. CZ.02.01.01/00/22\underline{\,\,\,}008/0004591, co-funded by the European Union.

\appendix

%\nocite{*}
\bibliographystyle{aipnum4-1}
\bibliography{aipsamp}% Produces the bibliography via BibTeX.

\end{document}